\documentclass{article}[12pt]

\usepackage[ascii]{inputenc}
\usepackage[T1]{fontenc}
\usepackage{geometry}
\usepackage{amsmath}
\usepackage{amssymb}
\usepackage[numbers,sort&compress]{natbib}
\usepackage[compact]{titlesec}
\usepackage{floatflt}
\usepackage{longtable}
\usepackage{wrapfig}
\usepackage{floatrow}
\usepackage{graphicx}
\usepackage{tabularx}
\usepackage{subcaption}
\usepackage{url}
\usepackage{color}
\usepackage{pdfpages}
\usepackage{enumitem}
\usepackage{authblk}

\usepackage{mathptmx}
\usepackage[english]{babel}
\usepackage{amsmath}
\usepackage{amssymb,amsfonts,textcomp}
\usepackage{array}
\usepackage{supertabular}
\usepackage{multirow}
\usepackage{hhline}
\usepackage{enumitem}
\usepackage{xspace}
\usepackage{upgreek}
\usepackage{microtype}
\usepackage{lineno}
\usepackage{stackrel}
\usepackage{accents}

\usepackage[colorlinks, urlcolor=blue, linkcolor=blue]{hyperref}
\makeatletter
\newcommand\arraybslash{\let\\\@arraycr}
\makeatother
\newcommand{\cpp}{C\kern-0.15ex{+}\kern-0.1ex{+}\xspace}

\newcommand{\FNAL}{Fermi National Accelerator Laboratory, Batavia, Illinois 60510, USA}
\newcommand{\LANL}{Los Alamos National Laboratory, New Mexico 87545, USA}
\newcommand{\Kavli}{Kavli Institute for Cosmological Physics at the University of Chicago, Chicago, Illinois 60637}
\newcommand{\UofCAA}{Department of Astronomy and Astrophysics, University of Chicago, Chicago, Illinois 60637}
\newcommand{\EFI}{Enrico Fermi Institute at the University of Chicago, Chicago, IL, 60637}
\newcommand{\ICERM}{Institute for Computational and Experimental Research in Mathematics, Brown University, Providence, RI, 02903}
\newcommand{\MIT}{Massachusetts Institute of Technology, Cambridge, Massachusetts 02139, USA}
\newcommand{\CMU}{McWilliams Center for Cosmology, Department of Physics, Carnegie Mellon University, Pittsburgh, PA 15213, USA}
\newcommand{\GOOGLE}{Google, Inc.}
\newcommand{\ADLER}{Department of Citizen Science, The Adler Planetarium, 1300 South Lake Shore Drive, Chicago, IL 60605, USA} 
\newcommand{\ZOONIVERSE}{Zooniverse}
\newcommand{\NORTHWESTERN}{Center for Interdisciplinary Exploration and Research in Astrophysics (CIERA), Northwestern University, 2145 Sheridan Road, Evanston, IL 60208, USA}
\newcommand{\CBPF}{Centro Brasileiro de Pesquisas F\'isicas, Rua Dr. Xavier Sigaud 150, CEP 22290-180, Rio de Janeiro, RJ, Brazil}

\begin{document}

\title{Response to NITRD, NCO, NSF Request for Information on Update to the 2016 National Artificial Intelligence Research and Development Strategic Plan}

\author[1]{J.~Amundson}
\author[1]{J.~Annis}
\author[2,3]{C.~Avestruz}
\author[1]{D.~Bowring}
\author[3]{J.~Caldeira}
\author[1]{G.~Cerati}
\author[2,4]{C.~Chang} 
\author[5]{S.~Dodelson}
\author[1]{D.~Elvira}
\author[5]{A.~Farahi}
\author[1]{K.~Genser}
\author[1]{L.~Gray}
\author[1]{O.~Gutsche}
\author[6]{P.~Harris}
\author[7]{J.~Kinney}
\author[1]{J.~B.~Kowalkowski}
\author[1]{R.~Kutschke}
\author[1]{S.~Mrenna}
\author[1,2,4]{B.~Nord}
\author[1]{A.~Para}
\author[1]{K.~Pedro}
\author[1]{G.~N.~Perdue}
\author[8]{A.~Scheinker}
\author[1]{P.~Spentzouris} 
\author[1]{J.~St. John}
\author[1]{N.~Tran}
\author[9]{S. Trivedi}
\author[10,11,12]{L.~Trouille}
\author[2]{W.~L.~K. Wu}
\author[13]{C.~R.~Bom}
\affil[1]{\FNAL}
\affil[2]{\Kavli}
\affil[3]{\EFI}           
\affil[4]{\UofCAA}        
\affil[5]{\CMU}           
\affil[6]{\MIT}
\affil[7]{\GOOGLE}        
\affil[8]{\LANL}         
\affil[9]{\ICERM}         
\affil[10]{\ADLER}         
\affil[11]{\ZOONIVERSE}   
\affil[12]{\NORTHWESTERN} 
\affil[13]{\CBPF}

\maketitle

\section{Introduction}

This document presents a response to the question of whether and how the National Artificial Intelligence Research and Development Strategic Plan (NAIRDSP) should be updated from the perspective of Fermilab, America's premier national laboratory for High Energy Physics (HEP). 
We believe the NAIRDSP should be extended in light of the rapid pace of development and innovation in the field of Artificial Intelligence (AI) since 2016, and present our recommendations below.
AI has profoundly impacted many areas of human life, promising to dramatically reshape society --- e.g., economy, education, science --- in the coming years.
We are still early in this process.
It is critical to invest now in this technology to ensure it is safe and deployed ethically.
Science and society both have a strong need for accuracy, efficiency, transparency, and accountability in algorithms, making investments in scientific AI particularly valuable.
Thus far the US has been a leader in AI technologies, and we believe as a national Laboratory it is crucial to help maintain and extend this leadership. 
Moreover, investments in AI will be important for maintaining US leadership in the physical sciences.

HEP is concerned with investigation of the most fundamental building blocks of matter and the forces that govern the universe. 
Like all the scientific disciplines, HEP will be transformed by the capabilities of AI for large-scale data analysis, pattern recognition, and anomaly detection. 
The HEP community contributes to the execution of a national strategy by advancing new technologies and by utilizing AI to solve some of the most difficult problems in science.
We are early adopters of powerful tools, like AI, because we must be --- to complete our mission we must push the best tools available to their limits and beyond.
Machine learning and deep learning have quickly become important tools of the trade for analysis of big data sets.
Given anticipated future data sizes and the complexity of problems, we have barely scratched the surface.

AI is more than algorithms. AI comprises data, sensors, computing platforms, and analysis techniques.
And HEP is well positioned to make unique contributions to each of these facets.
As a community, we work with data at enormous scale and develop novel, state of the art detector systems.
We readily embrace new computing platforms, and our focus on understanding bias and quantifying uncertainty will help with issues, such as algorithm explainability, cost-benefit analysis, and risk estimation in arenas outside of the sciences.
It is not enough to just design a new AI algorithm --- it must be connected to the real world, and we have the technical skills and use cases to be effective in this role.
Moreover, the data-driven algorithms of AI critically lack a statistical theoretical basis. 
It's crucial that we identify areas of AI theory where the physical sciences (data and methods) can inform AI tools and capabilities; this will create a positive feedback mechanism to the benefit of all parties. 
    
The applications should focus on accelerating science and on adding discovery methods for new science avenues. 
This should include the analysis of data, as well as the control of systems and instruments.
There should also be feedback of science domain knowledge and methods into developing the theory and tools of AI.
In order to facilitate the engagement of our community and the development of the necessary science applications, we believe it is essential to build partnerships across industries and academia in order to provide access to AI tools, develop the necessary infrastructure, and to share appropriate data sets for advancing these fields.
Discussed below are potential areas of research, development, application, and collaboration in AI that, from our perspective, will enable technological advancements to address scientific needs for HEP, as well as unique capabilities within the HEP community that can contribute to advancing AI.

\section{The Intersection of HEP and AI}


\subsection{Advanced Analysis for Discovery with Big Data}

High Energy Physics is deeply invested in the analysis of very large and complex data sets based on consistent underlying systems of meaning.
Indeed, the scale and complexity of modern data sets are exceeding the capacity of traditional data analysis algorithms and models. 
AI is evolving into an indispensable analysis tool for these data.
Moreover, we believe that HEP data has the potential to enable key insights into the theory and algorithms of AI itself.


\subsubsection{AI accelerates HEP science}
Modern implementations of AI algorithms are accelerating discovery for many critical tasks, like classification, measurement, and simulation \cite{Guest:2018yhq, Aurisano:2016jvx,PhysRevD.97.014021}.
Employing these AI algorithms has demonstrably sped up calculations (e.g. one million times for analysis of strong lenses \cite{2017Natur.548..555H}), and produced clear cost-savings (30\% increase in effective volume for neutrino flavor tagging with neural networks; \cite{PhysRevLett.118.231801,PhysRevD.98.032012}). 
These faster algorithms enable accelerated science, and in some cases they lead to paradigm shifts: in situations where it once took a human one day to analyze a single object, it may now take just a few seconds, pushing the frontier of discovery.

While AI is driving new analysis techniques, we find a suite of challenges in carrying these successes into the future and building on them. 
Most AI algorithms in use for science today require training sets assembled by humans, and they contain physics model biases. 
Nevertheless, high-precision hypothesis testing and modeling require accurate results that are free of bias.  
Even more, we must be able to interpret and explain the results coming from AI at least as well as we can from standard algorithms. 
But large AI models with millions of parameters typical of AI algorithms tend to lack interpretability. 
This because, in contrast with traditional models, individual parameters lack clear physical meaning, and measurements of error and uncertainty on these parameters differ in their statistical meaning from more traditional algorithms.
Modern AI algorithms are presently accelerating measurements, but they will be too slow for even many near-future data sets; we need to continue to improve training and inference time.
HEP experiments require large amounts of simulated data for calibration of instruments and mathematical models, but these simulations remain prohibitively slow. 
Finally, there are demonstrated successes in AI producing models of phenomena that are already known to us, but, how do we leverage the speed and flexibility of AI to make genuinely new discoveries --- outside the realm of current knowledge and beyond current training sets --- to produce new understanding of the Universe? 

A number of promising avenues can pave a path to the future of scientific discovery with AI. 
First, we must integrate statistical models with AI algorithms.
This will permit us to estimate and remove bias, and to produce statistically interpretable measures of uncertainty.
Next, we should incorporate elements of traditional physical models (whose parameters have physical meaning) into AI algorithms to improve interpretability. 
This also has the potential to improve algorithm speed: by leveraging well-understood physical laws through these physics parameters, it can limit the search space during optimization.
AI generative models can quickly produce simulated data sets, and we must learn how to deploy algorithms like these to more quickly produce simulations for calibrations and analyses \cite{PhysRevD.97.014021}. 
When data becomes sufficiently large, these algorithms will need to run on distributed systems \cite{2018arXiv180804728M}.

Beyond merely speeding up our analysis frameworks, AI has the potential to point us toward new theories of particle physics and cosmology. 
First, we discuss the connection between mathematical symmetries in AI algorithms and symmetries in nature.
HEP science is governed by symmetries of space, time, and energy. 
AI algorithms typically exploit only translational symmetry to optimize the search for patterns in data composed of flat images.
Recently, it has been shown that they can be generalized and implemented in a spherical context \cite{2018arXiv180609231K}.
This is more appropriate, for example, for data that has rotational and spherical symmetries, like the night sky that is on a spherical surface and which we see in upcoming large-scale cosmological experiments like CMB-S4 (Cosmic Microwave Background - Stage 4) \cite{2016arXiv161002743A} and LSST (Large Synoptic Survey Telescope) \cite{2009arXiv0912.0201L}. 
Also, consider the cases of force fields for molecular dynamics and properties of chemical compounds (REF; Kondor).
The symmetries here are yet more complex.
By matching the symmetry of the neural network architecture to the symmetry of the physical problem, we have a new opportunity for AI to learn the underlying physics in nature from data.

Second, we note the discovery potential that could be unleashed with `unsupervised' AI algorithms.
In these algorithms, there is no longer a training set from which the algorithm can learn patterns in the data before it sees the new target data; it must learn as it sees the data for the first time. 
For example, hybrid neural network and clustering algorithms that measure how similar each datum is from all others (REF).
Modern cosmological experiments, like LSST, will produce sufficiently large and complex data sets that searching for indications of new phenomena is prohibitively time consuming for standard non-AI algorithms. 
In these data sets, unsupervised algorithms could point us to outliers in the data, which could contain noise artifacts or new unexpected physical phenomena.
Finally, to effectively and efficiently tackle enormous data rates and volumes, while also preserving the capacity for serendipitous discovery, it is important to leverage the complementary strengths of humans and machines.
By keeping humans in the loop, these scenarios can embrace the speed of the algorithms while making them more accurate and interpretable. 
This has been demonstrated in citizen science efforts in galaxy classification (REFs; zooniverse)



\subsubsection{HEP brings insight to AI}

AI algorithms have been very successful on a number of standard data sets. 
For example, natural images of everyday objects (ImageNet \cite{imagenet_cvpr09,alexnet}) and handwritten digits (MNIST \cite{lecun-mnisthandwrittendigit-2010}) are canonical benchmarks for computer vision. 
Also, large-scale object detection, segmentation, and captioning data sets (COCO \cite{cocodset}) are available, as are standard environments for studying AI with games (OpenAI gym \cite{1606.01540}).

HEP has unique capabilities in data acquisition and simulation. 
HEP data is highly structured, self-consistent, and emerges from first principles.
We can produce very large, richly-labeled simulated datasets (that are easy to share) and as a field we specialize in understanding how they differ from nature. 
These data sets have meaning and explainability well beyond current natural image data sets, because they are developed with physical principals.
HEP data sets can provide rich, new benchmarks for all kinds of AI algorithms, from computer vision to reinforcement learning.

It is crucial to mitigate concentration of expertise into pools surrounding large private datasets in order to have broad participation across the economy. The 2016 strategy highlighted the need to develop shared public datasets and environments, and we should strongly extend those efforts. 
This supports workforce development and many other HEP initiatives that can result in opening public datasets for algorithm development - almost any project can include an open data component, but we can also build projects specifically around constructing platforms for dataset hosting and challenges. The democratization of frameworks and tools \cite{tensorflow2015-whitepaper,chollet2015keras,paszke2017automatic}, and the broad availability of commercial cloud computing services have opened the field to many participants - the difficult missing ingredient for initializing an AI research effort is no longer code, or access to modern compute infrastructure, it is access to large, richly structured datasets. HEP simulations can play an important role in solving this problem.

\subsection{Precise Control of Complex Systems}

Automation in our society is accelerating, and promises to provide new levels of efficiency, cost-savings, and improved standards of living.
Key examples are automation in transportation \cite{wiredcars} and industrial control \cite{deepcool}.
Self-driving cars and AI-augmented factory workforces are projected to improve safety and energy efficiency.
Enhanced facility management is already showing great energy-saving benefits: Consider Google's success with data center management. 
Over the past few years, Google has developed algorithms which leverage AI to recommend efficient ways to cool their data centers \cite{googlemldatacenter,deepcool,mitgoogledatacenter}. 
The resulting cost savings were significant enough that Google has automated the implementation of these recommendations, reducing power consumption by 40 percent and freeing up engineers to focus on more interesting challenges.

For HEP science, we must manage some of the most complex systems in the world.
For example, particle accelerators have thousands of elements to control and precisely tune within in very tight spatial and timing tolerances. 
In cosmology, telescope arrays scan the sky for highly diverse targets, and optimizing movements of the telescope is critical (REFs).
Nearly every experiment at national labs is dependent on quality operation of the accelerator complex. 
If automated control based on deep reinforcement learning \cite{suttonrl,Mousavi2018DeepRL}, and other strategies, is able to reduce downtime of machines, and free operators from mundane burdens to focus on more demanding and innovative work, the entire program benefits.
It is important for long-term efficiency and cost-effectiveness to begin automating the operation and control of these systems. 
We must use extremely fast systems for this, necessitating the development of real-time AI-based controls.



Within HEP, accelerator control proposals to date have considered only the optimization and automation of a single small system, e.g. \cite{Edelen:2016jgh}, but the real power of controls will materialize in the operation of large networks.
Modern reinforcement learning research is focused on hierarchical problem solving -- moving from global strategies all the way down into highly localized execution spaces \cite{DBLP:journals/corr/abs-1804-00810}.
Linking together multiple control systems across an accelerator complex into a larger network with greater responsibility and more opportunity for optimization is the end game, but execution of such a complex vision is a many-year project.
We need to do a great deal of research to make such a system possible, and even more to make it safe and explainable.
Furthermore, the physical complexity of accelerator systems and the huge number of difficult to learn constraints (for an exploratory algorithm), means that it will be critical to combine deep domain expertise with algorithmic research in order to build a functioning system.

One crucial challenge to the automatic control and optimization of large, complex systems, such as particle accelerators, is that they have many coupled time-varying components and the beam itself is time-varying and only partially observable. At best various 1D or 2D projections of the entire 6D phase space of the beam (x,x',y,y',z,$\Delta$E) are observable and usually with invasive/destructive methods. The fact that both accelerator components and the beams being input into them are varying with time limits the accuracy and ability of model-based methods. Even if an extremely deep neural network could be trained, using extremely large amounts of data, to precisely map desired beam characteristics (such as bunch length, energy spread, etc) to accelerator parameter settings, this network would only be accurate until the phase space of the injected beam and the accelerator components themselves drift from their original values. On the other hand, there are many model-independent approaches to optimization and control of uncertain and time-varying systems, one such example being developed at Los Alamos National Laboratory is Extremum Seeking (ES) \cite{ES_beam,ES,ES2}. While ES is able to simultaneously tune many parameters of uncertain and time-varying systems in a completely model-independent way, it is susceptible to getting stuck in a local minimum, as are all adaptive/model-independent feedback-based approaches. We believe that it is important to combine the two fields, AI and adaptive feedback, in order to enable adaptive machine learning approaches, in which model-based approacehs, such as neural networks provide instant, global approximations for parameter settings, based on a knowledge of the general features of large systems, and then local, model-independent feedback-based approaches zoom in on and track the optimal, uncertain time-varying parameters. 

In recent work, we performed a preliminary demonstration of such an approach, for the automatic control of 6 parameters of a particle accelerator in order to precisely control the longitudinal phase space (energy vs position) of a charged particle beam at femtosecond resolution \cite{PhysRevLett.121.044801} in a completely model-independent fashion, based only on images of the beam's longitudinal phase space and their comparison with a desired target distribution. The feedback-based method was able to tune, but did not match accurately and got stuck in a local minimum if it was initialized too far away from the optimal settings in the large parameter space. We then taught a neural network to directly map phase space distributions to accelerator settings. The neural network could give an approximation of the correct settings, but could not provide the exact required phase space becuase of both time-variation of the system and due to inerpolation as carried out by all learning networks. Finally, by combining the two methods, we were able to first get within a local neighborhood of the correct parameter settings by using the neural network, after which the adaptive feedback was able to zoom in and perfectly tune the beam to a desired phase space.





\subsection{AI at the Edge: Fast, Intelligent Sensors}
%
%



Many real-world problems require direct interaction between the computational resource and the physical environment. 
Sensor application complexity and integration are rising, and real-time collection and reaction times are declining. 
For example, in industrial settings, stringent requirements on sensitivity, safety, proximity, speed, and accuracy necessitate automation.
Sensors in these settings are not only numerous, but also diverse and sophisticated, ranging from high-resolution CCD detectors to high-speed particle detectors. 
This multiplicity of sensors can easily exceed what a person or a traditional computing system can accomplish in achieving the overall mission of such a facility.  

AI promises to address computing problems in this setting through a number of avenues: fast conversion of sensor signals to actionable information (e.g. classification or measurement of physical quantities); the intelligent synthesizing of information; and the routing of these data to larger global decision support and data analysis systems.
For AI systems to perform nearly any of these interactive functions, it will require real-time interaction with the environment --- to be embedded at the point of sensing. 
Systems that include large-scale sensor read-out and processing naturally have extremely heterogeneous compute environments --- utilizing the best computing elements for the job, accounting for factors like energy use and robustness against temperature, dirt, radiation, interference, and shock.  
AI promises to perform well (speed and accuracy) under these conditions, embedded directly in the problem domain. 
  
DOE National Labs, such as Fermilab, are well-situated to utilize AI because of their experience with the construction and operation of large-scale data acquisition and distributed control systems.  In addition, all necessary features for such interactions are present within their experimental facilities (the detection apparatus, as well as the equipment and buildings),  which will benefit immensely from the inclusion of these new AI capabilities at the edge.  To utilize these new technologies, new techniques and tools will be necessary to handle increases in data rates.
The larger computing facilities where sensor data is collected and processed will also require real-time analysis to form results and produce accurate reactions.  
For a large science experiment, this means ensuring that data is understood and correct.  
For a science collaboration, this means incorporation of newly collected data to form scientific results. 
Natrional Labs are in an excellence position for contributions because of their knowledge of ``big data'' science and algorithms for data analysis that can benefit AI.

Specialized compute hardware --- such as FPGAs, ASICs, and systems-on-a-chip --- comprise the base layers of the systems described above.
These components can (and will) incorporate AI inference to particle accelerator processing, performing the time-critical functions --- doing so at low power consumption rates and near to the collection and data transfer points.  
In the future, the higher layers of these edge-enabled systems will contain GPUs and specialized AI chips, such as TPUs (Google; REFs) and Neutral Engines (Apple; REFs).  
FPGAs are being worked into the higher-level components because of their ability to be dynamically reconfigured and handle large degrees of parallelism. 

Because of the experience developed over many decades with reconfigurable and special-purpose hardware, Fermilab and other National Labs are ideally suited to contribute to the construction of software (and firmware) algorithms to increase their utility.
An example of the need for very fast real-time data acquisition and handling is the LHC trigger systems at the front of the detectors, which must react in 100s of nanoseconds.  
Continuous read-out detectors, such as those present in neutrino experiments, need data acquisition completed in milliseconds.  
Many applications of future AI systems will involve analysis of real-time image data collected by cameras or other detectors, and in many cases these data may involve time evolution (``movies'') requiring processing at a scale far beyond the capabilities of humans.
The 50-picosecond time resolution of the dedicated cameras in modern smartphones, for example, enables robust and reliable face recognition through 3-D true depth imaging.  
High-resolution and -depth-sensing cameras may allow future AI-enabled systems to perform complicated tasks exceeding that of human capabilities, and can greatly simplify the computational complexity to improve the reliability of AI systems. 
Construction of large cameras with high spatial resolution and picosecond time resolution is well within current technical capabilities. 
The biggest challenge is in developing the integrated electronics systems to interface such cameras to the AI.

Exploration into using industry tools for AI have already begun for some data analysis problems to address real-time aspects of data processing. 
Google worked with the NASA to determine if machine learning could accelerate the search for exoplanets.  
This project, using Google's AutoML service and cloud TPUs, reduced single retrieval times from several days at 94 percent accuracy down to seconds, while increasing accuracy to 96 percent (REF).
These efficiency gains will allow NASA to now convert a serialized exploration into one that can run hundreds of retrievals simultaneously on the cloud. 
LSST is looking at use of Jupyterlab coupled with Kubernetes to aid in real-time interactive analyses (REF?).  
\subsection{Quantum Technology}
Investing in new hardware platforms and paradigms is an important part of long-term support for AI. In particular, investments in quantum information science (QIS) will play an important role in the long-term impact of AI. Quantum technology is a key component of a comprehensive response to the post-Moore's law era of computing \cite{postmoorequantum}. 
Furthermore, quantum sensors will enable entirely new programs for fundamental science.
It is critical to assume and maintain leadership in this technology and its applications.

Fermilab is seeking to leverage AI to help advance quantum science.
We are poised to become a leader in QIS over the next five years.
AI algorithms provide a powerful and efficient mechanism for tuning complex quantum systems in regimes where very fast times to solution are critical \cite{Heeres2017}.
The current state of the art for qubit coherence is tens of microseconds, so the time available to work with these systems is very short.

Machine learning can help control and optimize a quantum computer. The applications of quantum information science rely on optimal control of quantum devices to realize fast and high-fidelity state preparation, gate operations and readout \cite{ZHANG20171}. Designing these optimal control protocols involves the optimization of highly complex systems and hence is extremely challenging. Artificial intelligence is a powerful tool for the protocol design process to achieve performance beyond human capacity.

In terms of AI algorithms themselves, classical computers will dominate AI in the near and medium term, but over the longer time horizon of five to ten years from now, quantum advantages will greatly expand the reach and power of AI for certain problems.
Recent algorithmic advances in quantum machine learning, e.g. \cite{2018arXiv180609729V} and successful proof-of-principle algorithmic tests, e.g. \cite{Mott2017} have shown that quantum machine learning will eventually be able to leverage neural network connectivity beyond what is available to classical computers.
It is critical to establish leadership in a technology that may come to dominate AI applications.

We need to develop a quantum information and AI-aware workforce.
The convergence of quantum science and AI at Fermilab make this an ideal training ground for integrating these technologies with difficult, real-world applications supported by the right combination of expertise and infrastructure.

\subsection{Workforce Development}

Workforce development is one of the most crucial challenges posed by AI technologies.
AI has the potential to be a powerful engine of job creation and economic growth, as well as to improve the efficiency and safety for careers across almost all job sectors, including scientists.
Properly educating and training America's AI-enabled workforce will require academia, industry, and government working in tandem as an effective team.

We need to train not only the next generation of professional scientists to use these tools for research, but all citizens so they can become familiar and adept in the application of these tools in their careers.
Google has established programs in this area (REF), and a number of universities have established new majors (REF; CMU) or institutes of research and education (REF; MIT). 
National laboratories have significant experience in developing workshops and schools for training scientists on the job, for example well-established theory and accelerators schools (REFs). 
National labs can play a key role in that effort as a natural collaborative facilitator and nexus for these groups to partner on education and training ventures.  Fermilab, in particular, with its open access policy and partnerships with tech industry is well positioned to facilatate such activities.

\section{Organizational and Assessment Considerations}
\label{sec:orgconsiderations}

Organization and assessments are important components of initiatives and projects in general. 
Evaluating the degree to which Federal investments are successful is crucial for continuing support, as well as defining future R\& D directions for these programs. 
For the development of AI algorithms and applications for HEP, it is important to take into account that the field is in its early days.
AI has the potential for transformative impact on discovery science, and science is poised to provide insight and unique data sets for AI algorithm development.
It will be crucial to maintain US leadership in the development and deployment of AI, and that will require bringing all of our technical resources into proper alignment and fully utilizing the individual strengths of each field.

The broad applicability, uniqueness of scientific data, and need for algorithm theory development all necessitate partnerships between industry, National Labs and universities that have such capabilities. 
Given that algorithm experts are spread out among labs, academia and industry, a mix of such institutions is needed to carry out AI algorithm R\&D. The National Labs are an ideal platform for cooperation between academia, industry, and government.
HEP has much to offer to this enterprise in terms of large and richly detailed public datasets, a deep understanding of how to use simulation in real-world systems, cutting edge requirements for low-latency applications, and an important role in the educational and workforce development infrastructure of the country.
Labs feature the organization and infrastructure to manage large projects and have strong network effects throughout science in terms of drawing researchers together and serving as a locus for a wide variety of efforts. 
Given the broad, multi-disciplinary approach needed to successfully make progress in AI, it is natural to consider them as the hubs for such investment.

DOE National Labs are also well-situated to play a role as clearinghouses for data and algorithms.  In particular, HEP science requires the generation, storage, and curation of massive, complex data sets, and Labs such as Fermilab have cross-cutting resources, programs, and infrastructure that could be leveraged to share these data sets with the broader research community.

National laboratory campuses are designed to provide facilities for experimentation and computing. These facilities could be coupled to leadership-class compute facilities which provide massively large-scale GPU-enabled machines with thousands of nodes for production-scale work.  We envision a powerful ecosystem where utilization of small-scale resources (Labs, academia) is coordinated with utilization of large-scale resources of similar systems (Labs with leadership-class facilities). 
This would expose personnel at labs and nearby institutional partners to both the hardware and software environments of this computing infrastructure, giving them time to experiment and prepare for deployment of algorithms on the production-scale machines. 
This facilitates forward-thinking about hardware capabilities and the opportunity to get familiar with the devices for when they're deployed more broadly. 
This also provides complementary access and spreads out expertise, experience, and insight for next-generation devices.
Yet broader access to these AI-enabled computers may be facilitated with cloud infrastructure that is built to present a common portal for submission and management of workflows to the heterogeneous computing resources. 
Cloud capabilities will be especially important for fast development and testing of algorithms, and for unimpeded access to compute resources, which we believe will be essential for HEP science.  Fermilab has already developed and deployed infrastructure that allows access to cloud, leadership-class, and on-campus resources through a common interface (HEPCloud (REF)).   

To summarize, the frontier of scientific discovery is algorithmic in nature, and national labs can provide key resources and expertise as a nexus for research, collaboration, innovation, and workforce development to realize this future.

\newpage

\bibliographystyle{IEEEtran}
\bibliography{main}

\end{document}